\documentstyle[preprint,floats,prd,aps,epsfig]{revtex}

\preprint{\tighten\vbox{\hbox{\hfil CLNS 97/1526}
                        \hbox{\hfil CLEO 97-28}
}}
\begin{document}

\title {Improved Measurement of the Pseudoscalar Decay Constant $f_{D_s}$}

\author{CLEO Collaboration}
\date{\today}

\maketitle
\tighten
\begin{abstract}

We present a new determination of $f_{D_s}$ using 5 million $e^+e^-\to c
\overline{c}$ events obtained with the CLEO II detector.
Our value is derived from our new measured ratio
$\Gamma(D_s^+\to \mu^+\nu)/\Gamma(D_s^+\to \phi\pi^+)=0.173\pm 0.021 \pm 
0.031$. Using ${\cal B}(D_s^+\to\phi\pi^+) = (3.6 \pm 0.9)$\%, we extract
 $f_{D_s} = (280\pm 17 \pm 25 \pm 34 ){\rm ~MeV}$. We compare this result with
 various model calculations.
 
 PACS number(s): 1320.Fc, 13.20.-v
\end{abstract}
\newpage
{
\renewcommand{\thefootnote}{\fnsymbol{footnote}}

\begin{center}
M.~Chadha,$^{1}$ S.~Chan,$^{1}$ G.~Eigen,$^{1}$
J.~S.~Miller,$^{1}$ C.~O'Grady,$^{1}$ M.~Schmidtler,$^{1}$
J.~Urheim,$^{1}$ A.~J.~Weinstein,$^{1}$ F.~W\"{u}rthwein,$^{1}$
D.~W.~Bliss,$^{2}$ G.~Masek,$^{2}$
% ...
\end{center}
 
\small
\begin{center}
$^{1}${California Institute of Technology, Pasadena, California 91125}\\
$^{2}${University of California, San Diego, La Jolla, California 92093}\\
% ...
\end{center}
 
\setcounter{footnote}{0}
}
 
\newpage

\section{Introduction}

Measuring purely leptonic decays of heavy mesons allows the
determination of  meson decay constants, which connect measured quantities,
such as the $B\bar{B}$  mixing ratio, to CKM matrix elements.
 Currently, it is not possible to determine $f_B$ experimentally from 
leptonic $B$ decays, so theoretical calculations of $f_B$ must be used.
Measurements of the Cabibbo-favored pseudoscalar decay
constants such as $f_{D_s}$ provide a check on these calculations and
help discriminate among different models.

The decay rate for $D_s^+$ is given by \cite{Formula1} \cite{chargeconj}
\begin{equation}
\Gamma(D_s^+\to l^+\nu) = {{G_F^2}\over 8\pi}f_{D_s}^2m_l^2M_{D_s}
\left(1-{m_l^2\over M_{D_s}^2}\right)^2 \left|V_{cs}\right|^2~~~,
\label{equ_rate}
\end{equation}
where $M_{D_s}$ is the $D_s$ mass,
$m_l$ is the mass of the final state lepton, 
$V_{cs}$ is a CKM matrix element equal to 0.974 \cite{PDG}, and 
$G_F$ is the Fermi coupling constant.
Various theoretical predictions of  $f_{D_s}$ range from 190 MeV to 
350 MeV.
Because of helicity suppression, the electron mode $D_s^+ \to e^+\nu$
has a very small rate.
The relative widths
are $10:1:2\times 10^{-5}$ for the $\tau^+ \nu$, $\mu^+ \nu$
and $e^+ \nu$ final states, respectively.
Unfortunately the mode with the largest branching fraction,
 $\tau^+\nu$, has at least
 two neutrinos in the final state and is difficult to detect.

In a previous publication \cite{cleo}, CLEO reported the measurement  of
$f_{D_s} = (344\pm 37 \pm 52 \pm 42){\rm ~MeV}$, using the decay sequence
$D_s^{*+}\to\gamma D_s^+$, $D_s^+\to\mu^+\nu$. Three other groups have also
published the observation of $D_s^+\to\mu^+\nu$ and extracted values of
$f_{D_s}$. WA75 reported $f_{D_s}$ as $(232 \pm 45 \pm 20 \pm 48)$~MeV  using
muons from  $D_s^+$ leptonic decays seen in emulsions \cite{Bullshit};  BES
measured a value of  $(430 ^{+150}_{-130} \pm 40)$~MeV by fully
reconstructing $D_s^+$  mesons close to the  production threshold in $e^+e^-$
collisions \cite{Bes}; and E653 extracted a value of ($194\pm 35\pm 20 \pm 14$)
MeV from one prong decays into muons seen in an  emulsion target \cite{E653}.

In this paper we describe an improved CLEO analysis. We use  a
sample of about 5 million $e^+ e^- \to c\bar{c}$ events collected with the CLEO
II detector \cite{CIId} at CESR. The integrated luminosity is 4.79 fb$^{-1}$ at
the $\Upsilon (4S)$ resonance or at energies just below. This paper supersedes
our previous result which was based on a subset of the current data with 2.13
fb$^{-1}$. The improvements include a better analysis
algorithm, more data, more precise measurements of the lepton fakes, and
reduced systematic uncertainties.

\section{Analysis Method}
\subsection{Overview}
 
The analysis reported in this paper is based on procedures developed for the
previous CLEO II measurement of $f_{D_s}$ \cite{cleo}.  We search for the decay
chain $D_s^{*+} \to \gamma D_s^+$,   $D_s^+\to\mu^+\nu$. The photon from the
$D_s^{*+}$ decay and the muon from the $D_s^+\to\mu^+\nu$ decay are measured
directly, while the neutrino is measured indirectly  by using the
near-hermeticity of the  CLEO II detector to determine missing momentum and
energy. Using the missing momentum as the neutrino momentum,  we look for a
signal in  the mass difference
\begin{equation}
\Delta M = M(\gamma\mu^+\nu) - M(\mu^+\nu),
\label{equ_DeltaM}
\end{equation}
so that the relatively large errors from the
missing momentum calculation will mostly cancel.

To study the $\Delta M$ signal and background shapes
 and to evaluate the effectiveness of our Monte Carlo efficiency 
simulation, we also collect a data sample of similar 
topology, $D^{*o}\to \gamma D^o$, $D^o \to K^-\pi^+$. We  
treat these fully reconstructed data events as $D_s^+\to\mu^+\nu$ decays
by removing the measurements of the $\pi^+$
from both the tracking chambers and the calorimeter
to simulate the $\nu$, and by ``identifying" the $K^-$ as a muon. Our aim
here is to compare the Monte Carlo simulation of these $D^{*o}$ decays
with what we obtain from the data.

Another useful event sample consists of the decay sequence
$D^{*+}\to \pi^+ D^o$, $D^o\to K^-\pi^+$, since this sample has
relatively high statistics and negligible background.
We use these events to study the missing energy and momentum
measurements by eliminating
the measurements of the fast $\pi^+$ from the $D^o$ decay from both the
tracking chambers and calorimeter to simulate the neutrino, and call
the $K^-$ a muon. 

\subsection{Background}

 There are several potential sources of background for this measurement.
The real physics backgrounds, such as semileptonic decays,
are almost identical in muon and electron final states
because of lepton universality.
For the leptonic $D_s^+$ decay, however,
the electronic width is negligible in comparison to the muonic
width.
Thus, performing the identical analysis except for
selecting electrons rather than muons gives us a quantitative
measurement of the background level due to real leptons. 
$D_s^+\to\mu^+\nu$ and $D^+\to\mu^+\nu$ are the only physics processes
that produce
significantly more primary muons than electrons with momenta above 2~GeV/{\it c}
in continuum $e^+e^-$ annihilations in the $\Upsilon(4S)$ energy region.
$D^+\to\mu^+\nu$ decay background in our sample is highly suppressed 
by the CKM angle (Eq. \ref{equ_rate}), and by
the small $D^{*+} \to \gamma D^+$ branching
ratio, ($1.4\pm 0.5\pm 0.6$)\%  \cite{Randomg}.

Another source of background results from the misidentification of hadrons as
muons (fakes).  Since muon identification in CLEO II has larger fake rates than
electron identification, we need to consider the excess fakes in the muon
sample relative to the electron sample. To determine the hadron-induced muon
and electron  fake background contributions, we multiply the  $\Delta M$
distribution of all tracks, excluding identified leptons, by an effective
hadron-to-lepton fake rate, measured with tagged hadronic track samples.  The
detailed analysis of this effective fake rate is described in Section
~\ref{background}.

After removing the above two components, all remaining events result from
either $D_s^{*+}\to\gamma D_s^+$, $D_s^+\to\mu\nu$ decays, or
from spurious combinations of random photons and
real $D_s^+\to\mu^+\nu$ and $D^+\to\mu^+\nu$ decays.
The shape of the latter component is determined
using the fully reconstructed $D^{*+}\to\pi^+ D^o, D^o\to K^-\pi^+$
data sample, and the normalization is determining by measuring the 
$D_s^{*+}/D_s^+$ production ratio. Subsequently, we will form a
single signal shape from these two signal components.

\subsection{Event Selection and Background Suppression}
\label{Event}

Most of the leptons from $B$ meson decays are removed by
requiring a minimum lepton momentum of 2.4~GeV/{\it c}, which
is 33\% efficient for $D_s^+\to\mu^+\nu$.
Leptons from $\tau^+\tau^-$ pairs, and other QED processes with low
multiplicity, are suppressed by requiring that the event either
has at least five well reconstructed
charged tracks, or at least three charged tracks accompanied by at least
six neutral energy clusters.
To suppress background from particles that escape detection
at large cos$~\theta$, where $\theta$ is the angle 
with respect to the beam axis, we require that the
angle between the missing momentum of the event and the beam axis, 
$\theta_{miss}$, does not point along the beam direction, specifically
$|{\rm cos}~\theta_{miss}|< 0.9$.

Muons are required to penetrate at least seven
interaction lengths of iron, and to have
$|{\rm cos}~\theta|<0.85$. The muon identification
efficiency, measured with
$e^+e^-\to \mu^+\mu^-\gamma$ events,
is (85$\pm$1)\% for muons above 2.4 GeV and is very flat in  momentum.
Electrons must have
an energy deposit in the electromagnetic calorimeter close to
the fitted track momentum, and a $dE/dx$ measurement in the main drift chamber
consistent with that expected for electrons.
The electron identification efficiency
for $|{\rm cos}~\theta|<0.85$, is
found by embedding tracks from radiative Bhabha events into hadronic events.
For electrons with momentum greater than 2.4 GeV, a  value
of (89$\pm$2)\% is used. 


To subtract the electron data from the muon data we need to have
a precise measure of the muon to electron
normalization. Detector material causes a difference between muons
and electrons, as electrons tend to radiate more. The correction factor
is estimated to lower the electron rate by 5\%: thus we assign a +5\%
increase in the electron sample due to this outer bremmstrahlung.
A Monte Carlo study shows that the main background contributions from real
leptons in
the $\Delta M$ distribution are  semileptonic $D$ decays, mostly $D\to
K\ell\nu$, $\pi\ell\nu$ and $\eta l\nu$. As a specific example of the near
equality of the muon and electron rates we made a detailed study of
the $D^+\to K^o\ell^+\nu$ decay.
A calculation of 
the different probabilities that a photon is emitted in the
decay (inner bremsstrahlung) for  $D^+ \to K^o l^+\nu$ was performed
according to the prescription of Atwood and Marciano \cite{rocky}. This
effect
raises the electron rate by +2.7\%. This inner bremsstrahlung
correction for the different semileptonic final states averages also to +2.7\%. 
We also correct for  
differences in muon and electron phase
space, which lowers the relative electron normalization
($-1.7\%$ for $D^+ \to K^o l^+\nu$ ).
Taking all of these sources into account, including the different
possible decay modes and the fact that the electron
detection efficiency is 4\% larger than the muon efficiency,
 we use a correction factor
of 1.01$\pm$0.03 to multiply the electron sample to account for the 
physics backgrounds and the identification efficiency difference.

Photons must be in the angular region
$|{\rm cos}~\theta|<0.71$.
We require a minimum energy
of 150 MeV, which is 78\% efficient for $D_s^{*+} \to \gamma D_s^+$ decay,
to eliminate backgrounds caused by the large number
of low energy photons.
Combinations  of two photons which have invariant masses within
two standard deviations of the $\pi^o$ mass are
eliminated. (The r.m.s. $\pi^o$ mass resolution is 5 MeV.)
We also insist that in the rest frame of the
$D_s^{*+}$ candidate, the cosine of the angle between the photon and the 
$D_s^{*+}$
direction in the lab be larger than $-0.7$. 
A small residual $b\to u l \nu$ background
is suppressed by requiring that
the thrust axis lines up with the $D_s^{*+}$ candidate momentum
so that the cosine of the
angle between them is greater than 0.975.

\subsection{Signal Shape and Efficiency}
\label{Signal}

To evaluate the neutrino four-vector we measure
the missing momentum and energy in only half of the event;
we divide the event into two hemispheres using
the thrust axis of the event.
The missing momentum $\vec{ p}_{miss}$
and energy $E_{miss}$ are calculated using only
energy and momentum measurements ($E_i, \vec{ p}_i$)
in the hemisphere
that contains the lepton (kaon).
We compute the energy sum assuming all tracks are pions, unless they
are positively identified as kaons, or protons by {\it dE/dx} measurement
in the drift chamber.
We define the missing momentum and energy as
\begin{eqnarray}
\vec{ p}_{miss}	&=	&\vec{
p}_{thrust}-\sum\vec{ p}_i~~~{\rm and}\nonumber \\
E_{miss}			&=	&E_{beam}-\sum E_i~~~,
\end{eqnarray}
where the direction of $\vec{ p}_{thrust}$ is given by the thrust
axis. The magnitude is
$p_{thrust}^2=E_{beam}^2-m_{jet}^2$,
where $E_{beam}$ is the beam energy and $m_{jet}$
is the average mass of a charm quark jet,
 measured to be 3.2 GeV using  our sample of fully reconstructed
  $D^{*+}$ events \cite{Dstarplus}. 
A $D_s^+$ candidate is selected by requiring 1.2 GeV $<M(\mu^+\nu)<$ 3.0 GeV,
and that 
the missing mass squared be consistent with a neutrino,
$|E_{miss}^2-p_{miss}^2|<$ 2 GeV$^2$, where the cut values are based on
studies using the $D^{*+}$ events. Furthermore, we
 also require $p_{miss}>0.8$ GeV/{\it c} to suppress backgrounds, since real
$D_s^+\to \mu^+\nu$ events must have some missing momentum. 
The $D_s^{*+}$ candidate momentum is required
to be above 2.4 GeV/{\it c}.
We find a factor of two increase in efficiency by using only one hemisphere
to determine the missing momentum relative to using the whole event.

Although the measurement errors on the muon and neutrino tend to cancel when
evaluating the mass difference in Eq.~\ref{equ_DeltaM}, the neutrino is poorly
enough measured  to cause a significant broadening of the resolution in
comparison with fully reconstructed $D_s^{*+}$ samples. Improvement is possible
by using the constraint that the muon and neutrino four-vectors must have the
$D_s$ invariant mass. Since the muon is much better measured than the neutrino,
we vary only the neutrino momentum relative to the selected muon. From
conservation of energy and momentum, we have \begin{equation} E_{D_s} = E_{\mu}
+ E_{\nu}  {~~~\rm and}\label{equ_energy} \end{equation} \begin{equation}
\vec{p}_{D_s} = \vec{p}_{\mu} + \vec{p}_{\nu}.
\label{equ_mom} \end{equation}

Squaring  Eq.~\ref{equ_mom} in the local coordinate frame defined by the muon
and the reconstructed neutrino, using Eq.~\ref{equ_energy} and rearranging
shows a relationship between $p_{\nu}$ and the cosine of the angle between the
muon and neutrino:
\begin{equation}
p_{\nu}=\left(m_{D_s}^2-m_{\mu}^2  \right)/\left(2E_{\mu}-2p_{\mu}\cos{\theta}
\right), {~~~\rm where~~ E_{\mu}=\sqrt{m_{\mu}^2+p_{\mu}^2}}.  \label{equ_cos}
\end{equation}

\begin{figure}[p]
\centerline{\epsfig{figure=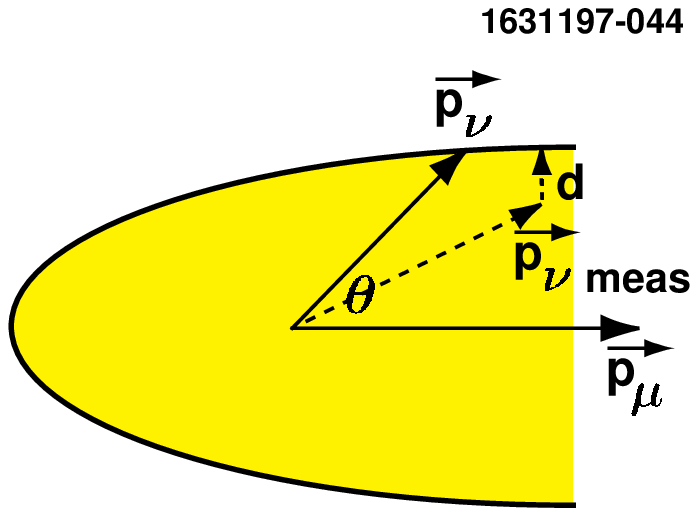,height=3in}}
\caption{ \label{Ds_const}The relationship between the muon and neutrino
momentum vectors and the constraint surface imposed by the $D_s$ invariant mass.
}
\end{figure}

Fig.~\ref{Ds_const} shows the constraint as a surface of revolution about the
muon momentum vector. We start by defining a plane by the vector cross product
of the measured muon and neutrino three-vectors, though the ``correct" solution may
lie outside this plane. We next find the minimum distance from the
measured neutrino momentum vector to the surface. Clearly, the new neutrino
momentum is the vector sum of the measured neutrino momentum
$\vec{p}_{\nu \rm{~meas}}$ and the distance in momentum space,
$\vec{\rm d}$, as is shown in
Fig.~\ref{Ds_const}. This procedure improves the $\Delta M$ resolution by about
30\%.   

We use Monte Carlo simulation to determine the $\Delta M$ signal shape
(Eq.~\ref{equ_DeltaM}) and to estimate our efficiency.
Since this analysis involves reconstructing a missing
neutrino, we are concerned that the Monte Carlo will not adequately simulate
the data. As a check we evaluate the accuracy of our simulation using our 
$D^{*o}\to\gamma D^o$, $D^o\to
K^-\pi^+$ sample, where we eliminate the $\pi^+$ to simulate the neutrino
and treat the $K^-$ as a muon.

\begin{figure}[p]
\centerline{\epsfig{figure=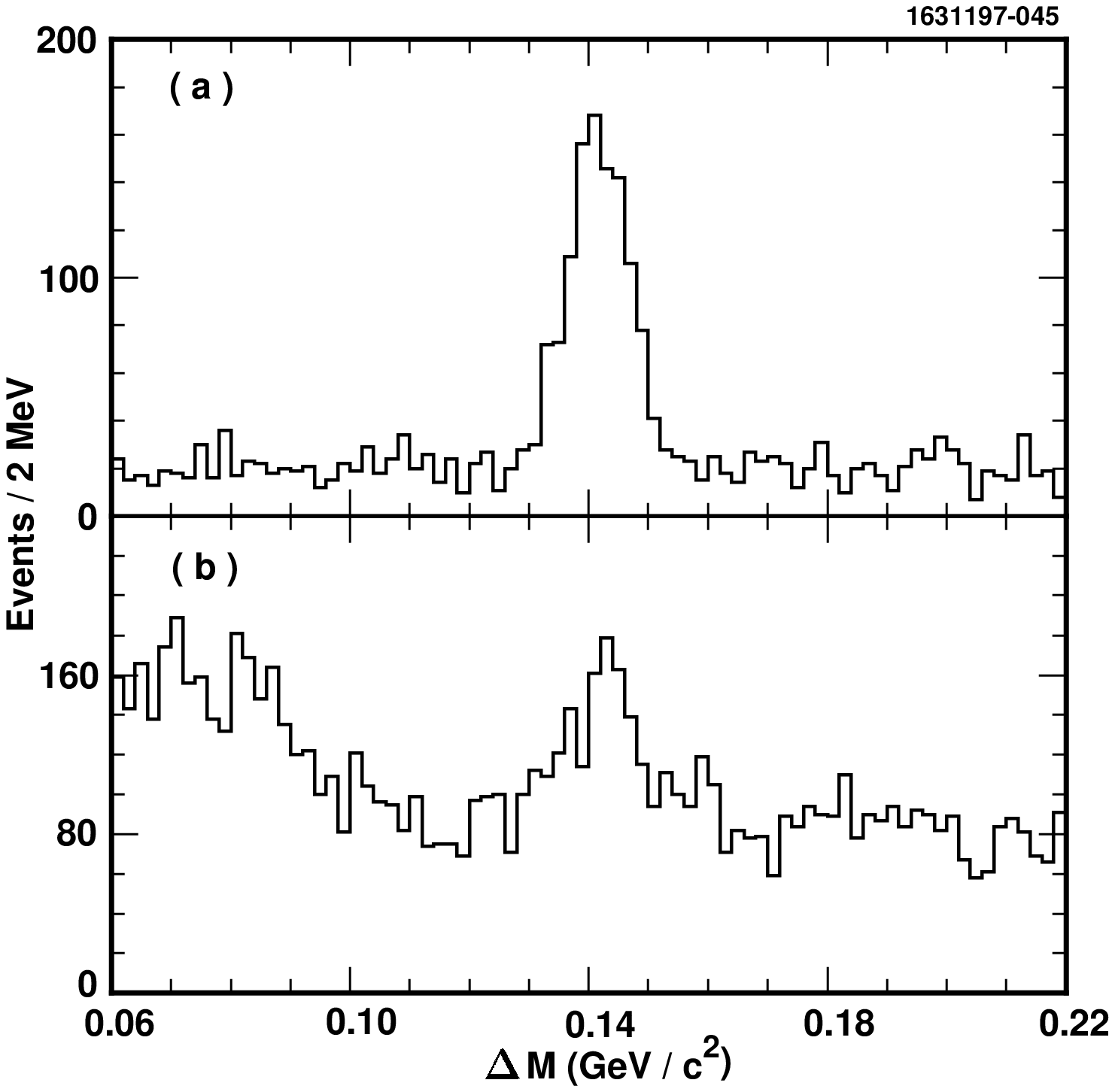,height=6in}}
\vspace{.5cm}
\caption{\label{Dsz} The $\Delta M=M(\gamma K^-\pi^+) -
M(K^-\pi^+)$ mass difference distributions for fully
reconstructed events of  $D^{*o}\to\gamma D^o$, $D^o\to K^-\pi^+$,
after a requirement  that the $K^-\pi^+$ mass be within 2.5 standard
deviations of the $D^o$ mass. (a) $D^{*o}$ Monte Carlo (b) data.}
\end{figure}

\begin{figure}[hb]
\centerline{\epsfig{figure=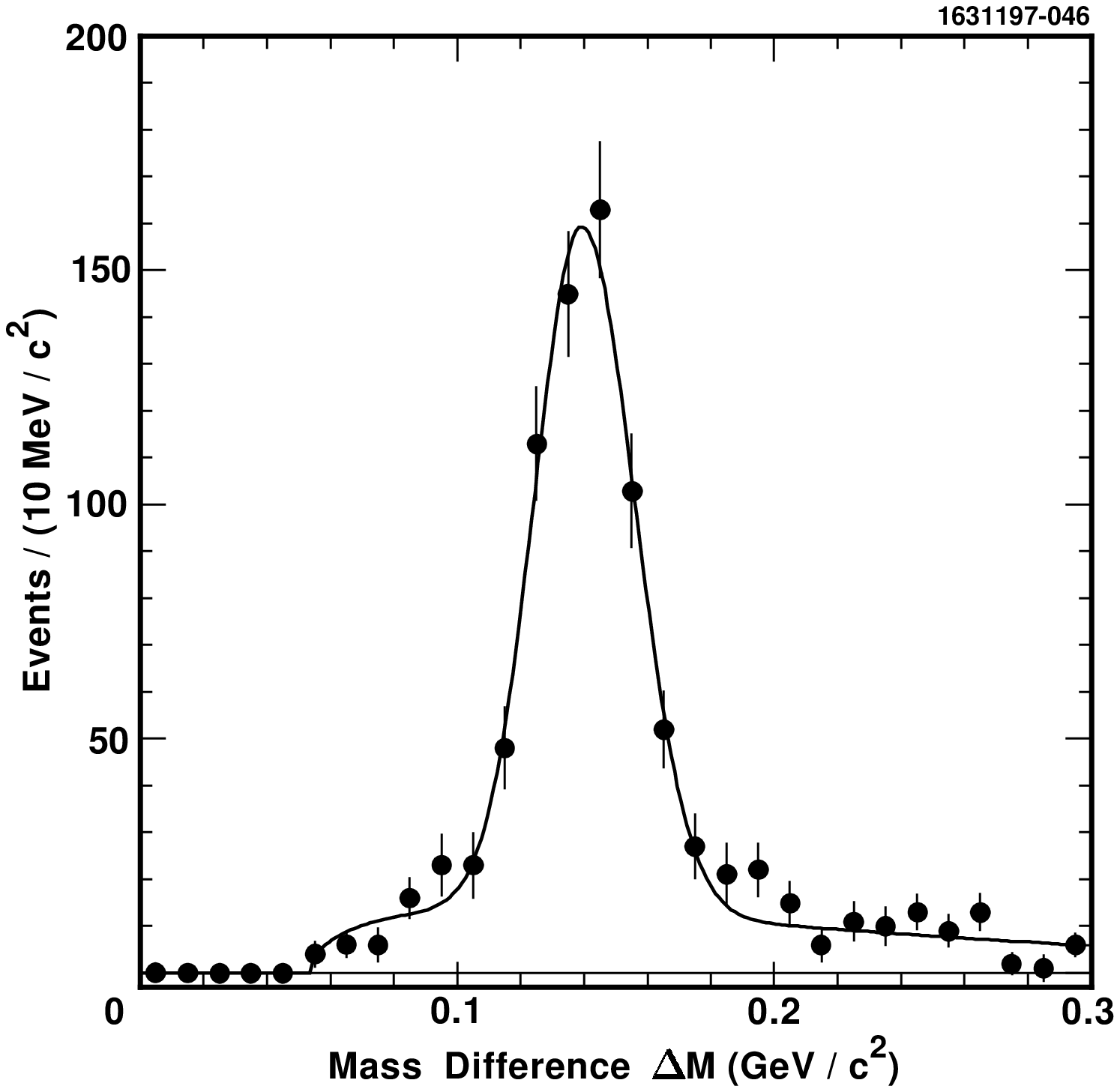,height=5.5in}}
\vspace{.5cm}
\caption{\label{Fsig} The $\Delta M=M(\gamma Kp_{miss}) -
M(Kp_{miss})$ mass difference distributions for
the simulated missing momentum analysis using Monte Carlo of 
 $D^{*o}\to\gamma D^o$, $D^o\to K^-\pi^+$. A sideband
subtraction to remove background in the initial $D^{*o}$ selection
has been applied. The curve and fitting procedure are described in the text.}
\end{figure}

We start with a $D^{*o}\to\gamma D^o$, $D^o\to K^-\pi^+$ Monte Carlo simulation. 
Fig.~\ref{Dsz}(a) shows the fully reconstructed mass difference
$\Delta M= M(\gamma K\pi) - M(K\pi)$ distribution after a cut on the $K^-\pi^+$ 
invariant mass of $\pm$30 MeV around the known $D^o$ mass (where the r.m.s 
resolution is 8 MeV). The kaon is required to have momentum greater than
2.4 GeV/{\it c}, which is the same cut as we use on the muon in the $D_s^+\to\mu^+\nu$
channel.
In the $\Delta M$
distribution there is a substantial signal but also
significant background, so a $\Delta M$ sideband subtraction is performed.
We use a bin-by-bin subtraction. The central value of the signal is 142 MeV and
the r.m.s. width is 5.5 MeV. The sidebands used are 114-126 MeV and 159-170
MeV.
After applying the additional background suppression cuts, described above, we
obtain the mass difference distribution $\Delta M = M(\gamma Kp_{miss}) -
M(Kp_{miss})$ shown in Fig.~\ref{Fsig}. There is a clear signal peak
associated with the photon and it is fitted to an asymmetric Gaussian with low
side and high side $\sigma$'s of 15 MeV and 16 MeV, respectively. 
The small flat component results from replacing the correct photon with 
another photon. 

The partial efficiency for neutrino detection only from Monte Carlo for a fully
reconstructed $D^{*o}$ event  with both the $D^{*o}$ and its kaon daughter
having a momentum greater than 2.4 GeV to appear in the signal peak after
neutrino reconstruction is found to be $\epsilon_{\nu}$ = (38.9$\pm$2.6)\%
\cite{denom}. 
The overall detection  efficiency for $D^{*o}\to D^o \gamma$, $D^o\to K^+
p_{miss}$ is $(4.8 \pm 0.3)\%$ \cite{correff}.

\begin{figure}[hb]
\vspace{-2.5cm}
\centerline{\epsfig{figure=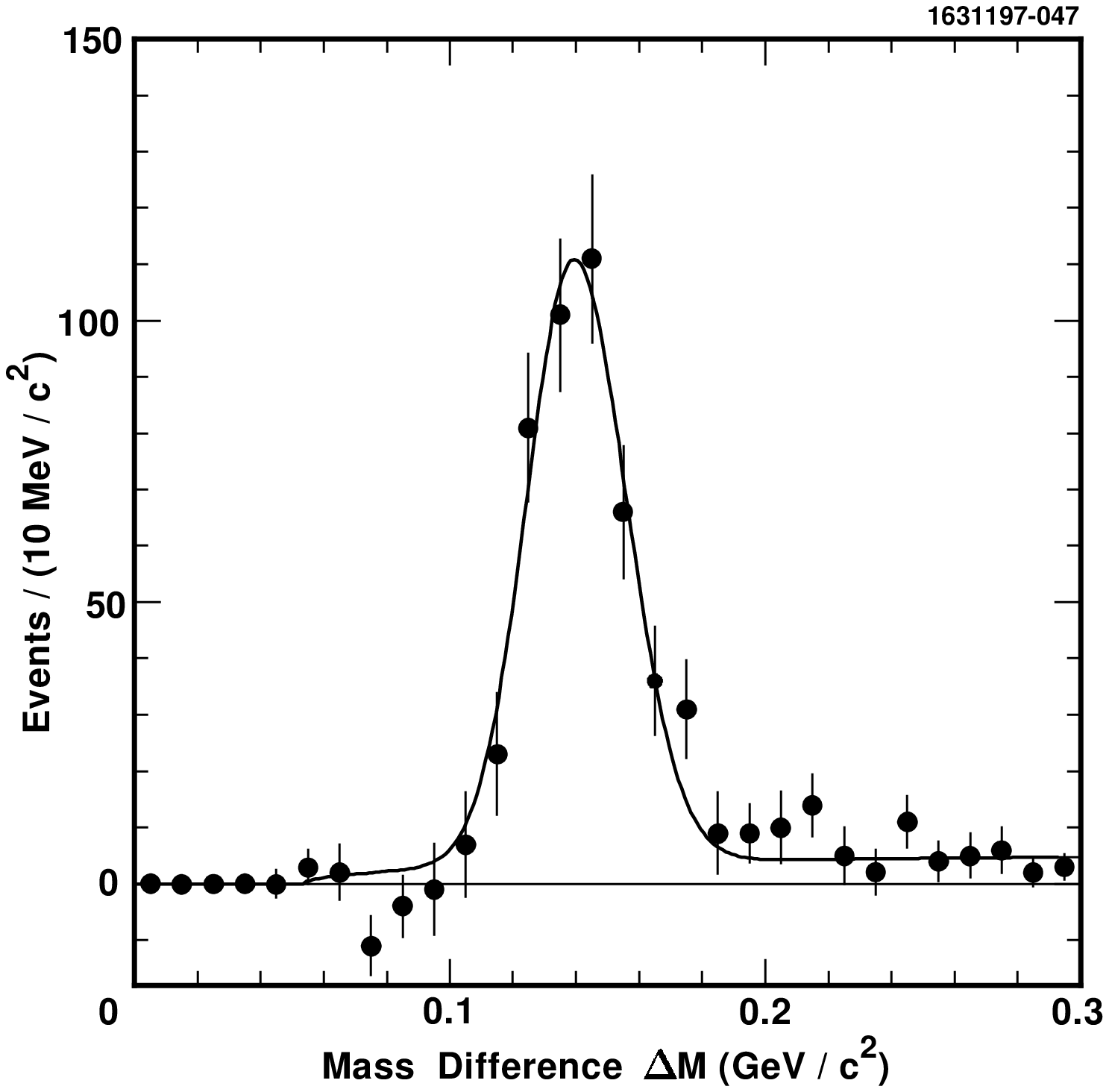,height=6in}}
\caption{\label{gd0mas} The $\Delta M=M(\gamma Kp_{miss}) -
M(Kp_{miss})$ mass difference distributions for
the simulated missing momentum analysis for
the $D^{*o}\to\gamma D^o$, $D^o\to K^-\pi^+$ real data. A sideband
subtraction to remove background in the initial $D^o$ selection
has been applied. The curve used is from the Monte Carlo signal
shape.}
\end{figure}

Next, we repeated the analysis described above for the
fully reconstructed 
$D^{*o}\to \gamma D^o$ data sample. The fully reconstructed $\Delta M$ distribution
is shown in Fig.~\ref{Dsz}(b). The $\Delta M$ distribution for the missing neutrino is shown
in Fig.~\ref{gd0mas} where the sideband subtraction again has been performed. 
The fitting parameters derived from the Monte Carlo signal shape
fit the data very well, with a $\chi^2$ of 23 for 27 degrees of freedom and a
confidence level of 69\%. 
The partial efficiency for neutrino detection oly of (38.5$\pm$3.7)\%
agrees well with Monte Carlo simulation.

\begin{figure}[hb]
\vspace{-2.5cm}
\centerline{\epsfig{figure=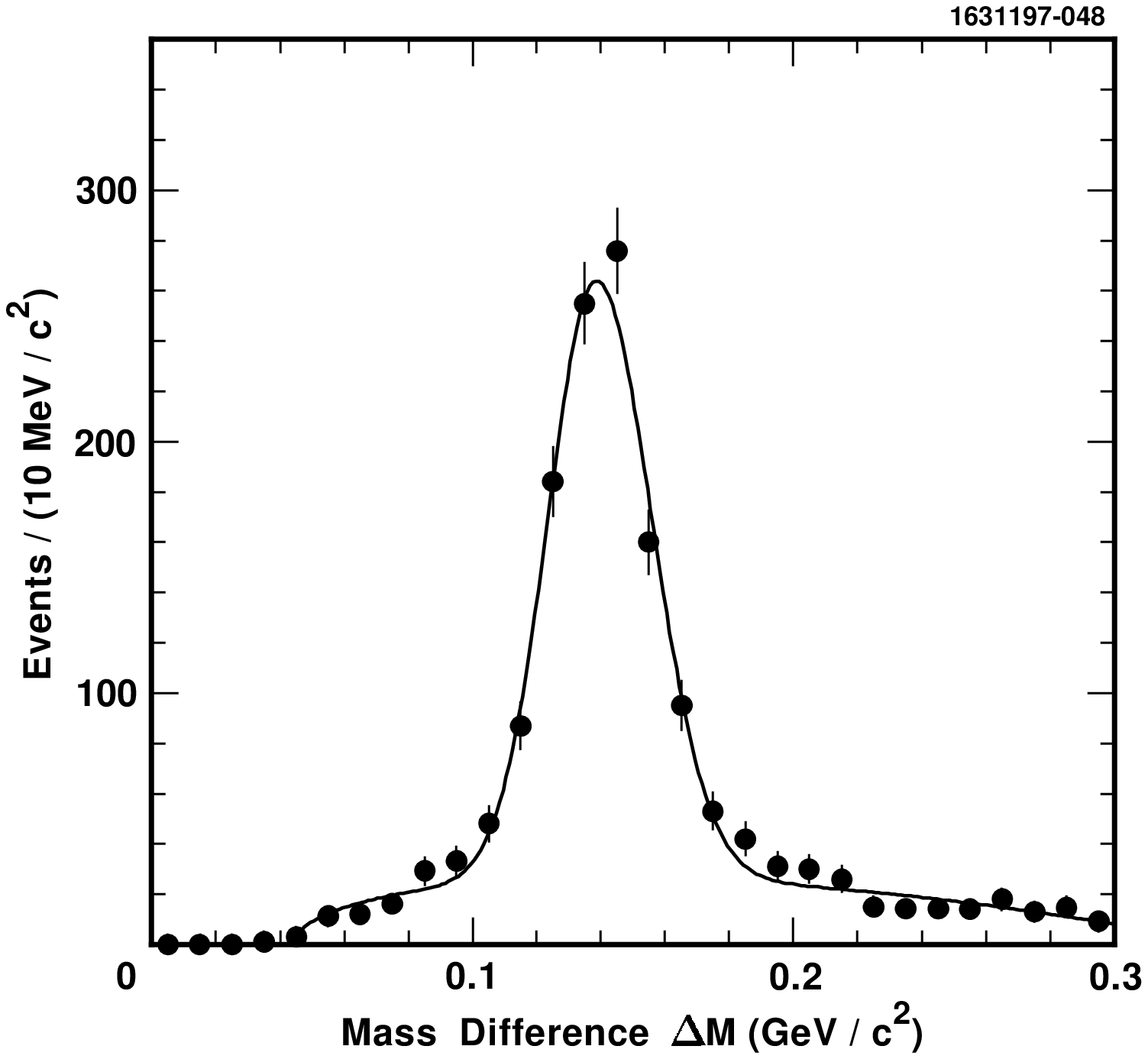,height=6in}}
\caption{\label{gdsmas} The $\Delta M=M(\gamma\mu p_{miss})-M(\mu p_{miss})$ 
mass difference distributions for
the missing momentum analysis for
the $D_s^{*+}\to\gamma D_s^+$, $D_s^+\to \mu^+\nu$  Monte Carlo.
The curve and fitting procedure are described in the text.}
\end{figure}

In principle the resolution and efficiency for $D_s^{*+}\to \gamma D_s^+$,
 $D_s^+\to
\mu^+\nu$ can be somewhat different from that for the $D^{*o}$ sample described above,
because of the different fragmentation with an $s$ quark rather than a $u$ quark.
Since our Monte Carlo simulation accurately describes the
$D^{*o}\to\gamma D^o$, $D^o\to K^-\pi^+$ process, we 
rely on it for our $D^*_s$ study. In Fig.~\ref{gdsmas} we show
the $\Delta M=M(\gamma\mu^+\nu)-M(\mu^+\nu)$ distribution from
Monte Carlo simulation. This distribution contains a Gaussian part
due to the signal, plus a background which occurs when the correct
photon from the $D_s^*$ decay is replaced with another random photon
in the event. We fit the histogram  
with an asymmetric Gaussian signal shape having low
side and high side $\sigma$'s of 15 MeV and 17 MeV, and
 the function 
$\sqrt{x-x_0}e^{-a(x-x_0)}$ to parameterize the random photon component,
where $x\equiv\Delta M$.
The Gaussian signal shape agrees well with the $D^{*o}$ Monte Carlo and data.
Using the Gaussian signal part only,
the overall efficiency is found
to be (4.2$\pm$0.3)\%, where the error includes the systematic effect
of the efficiency difference between 
data and Monte Carlo determined by the $D^{*o}$ sample.

\begin{figure}[hb]
\centerline{\epsfig{figure=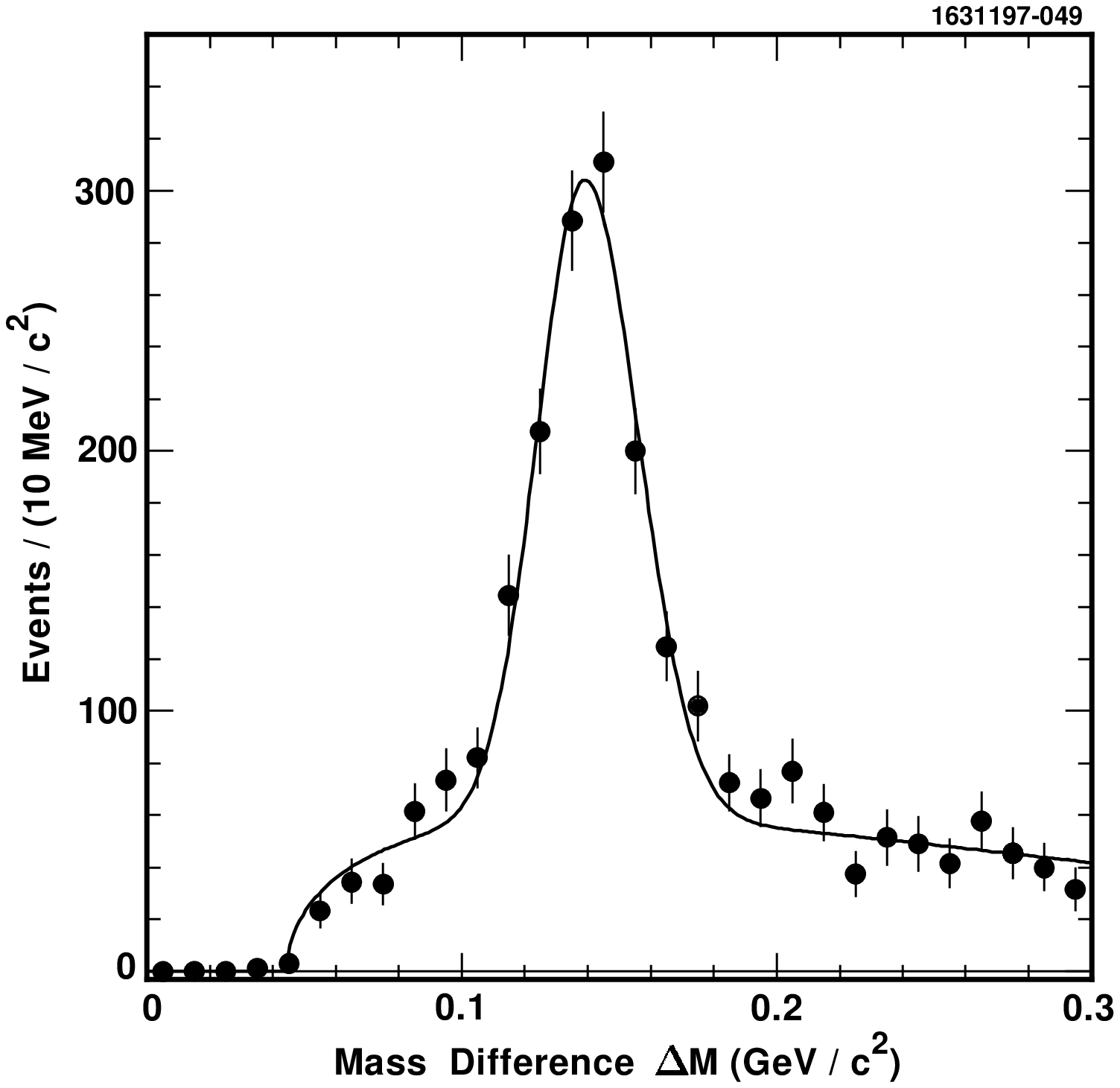,height=6in}}
\caption{\label{signal_flat_bkgd} The $D_s^{*+}\to\gamma D_s^+$,
 $D_s^+\to\mu^+\nu$ signal distribution plus random photon background as determined from the 
 signal Monte Carlo simulation combined with the
 $D^{*+}$ data sample analyzed for the missing $\nu$ as 
 $D^{*o}\to \gamma D^o$, $D^o\to K^- p_{miss}.$
 The curve is a fit using the functions described in the
 text.}
\end{figure}

An additional source of background in the $\Delta M$ distribution
comes from
 direct $D_s^+\to\mu^+\nu$ decays which pair with
a random photon to form a $D_s^{*+}$ candidate. These are in addition to
$D_s^{*+}$ events where the correct photon is replaced by another photon,
as mentioned above.
These two contributions are fixed relative to the direct $D_s^{*+} \to
\gamma D_s^+, D_s^+ \to \mu \nu $ signal using our measurement of
$D_s^{*+} / D_s^+$ production ratio above 2.4 GeV of 1.08$\pm$0.13 (see below).  
Thirdly, there is a small contribution from $D^+\to\mu^+\nu$ decays
combined with a random photon.
The shape in $\Delta M$ of all these contributions is modeled 
using the $D^{*+}\to\pi^+ D^o$ event sample, by combining the
$M(K p_{miss})$ candidates with random photons in the same event,
and fitting with the functional form 
$\sqrt{x-x_0}e^{-a(x-x_0)}$ to parameterize the total random photon component.
The distributions in Fig.~\ref{gdsmas} and the random photon
component function are summed using appropriate weights to produce the
expected  shape for the sum of the $D_s^{*+}\to \gamma D^+_s,
~D_s^+\to \mu^+\nu$ signal plus random photon background shown in 
Fig.~\ref{signal_flat_bkgd}.

\subsection{Measurement of the $D_s^+$ and $D_s^{*+}$ Rates}

In order to measure the relative rates of $D_s^*$ and $D_s$ production,
and the absolute level of $D_s$ production above 2.4 GeV/{\it c} we
use the $D_s^+\to \phi\pi^+$ decay mode. The $\phi$ is searched for in
the $K^+K^-$ decay mode. We require the photon from the
$D_s^{*+}$ decay to satisfy the same requirements as for the $\mu^+\nu$ final
state. The detection efficiency for the $\phi\pi^+$ decay mode is 22.3\%,
while for the $D_s^{*+}$ the efficiency is 9.4\% \cite{effcom}.

\begin{figure}[hb]
\centerline{\epsfig{figure=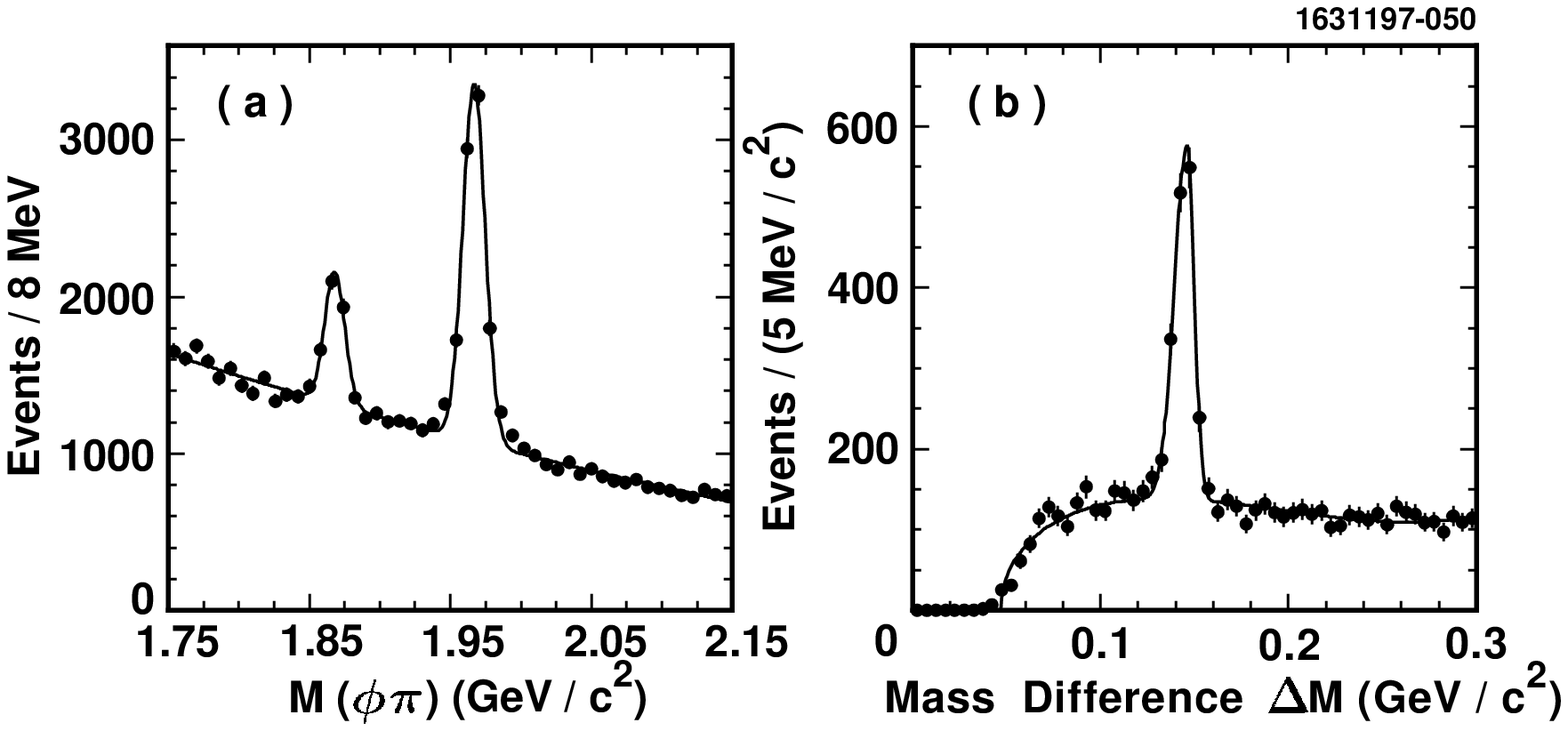,width=6in}}
\vspace{.5cm}
\caption{\label{phipi} (a) The $\phi\pi^+$ mass distribution and (b) the 
$\Delta M=M(\gamma\phi\pi^+)-M(\phi\pi^+)$ mass difference distribution
with the requirement that $\phi\pi^+$ mass is consistent with the 
known $D_s^+$ mass. The signal shapes are taken from Monte Carlo simulation.
The background shape in (a) is a second order polynominal, while
in (b) it is the sum of half-integer polynomials.}
\end{figure}

Fig.~\ref{phipi}(a) shows both the invariant mass of the $\phi\pi^+$. In (b),
we show $\Delta M = M(\gamma\phi\pi^+)
-M(\phi\pi^+)$ after requiring that the $\phi\pi^+$ mass be within 
$\pm$24 MeV of the $D_s^+$ mass.

Fitting the data to Gaussian signals shapes whose widths are determined by
Monte Carlo simulation we find 5728$\pm$123 $D_s^+\to\phi\pi^+$ events and
1256$\pm$54 $D_s^{*+}$ events. Taking into account the relative efficiencies we
determine that the ratio of $D_s^{*+}/D_s^+$ production is 1.08$\pm$0.13. This
number reflects the direct production of a vector charmed-strange meson
relative to the direct production of a pseudoscalar charmed-strange meson,
above 2.4 GeV/{\it c} \cite{goodVP}.

\section{Lepton Fake Background Calculation}
\label{background}

Even after strict lepton identification requirements have been applied, 
significant numbers of
hadron fakes still enter our signal region because of the abundance of
fast hadron tracks.
To properly account for the hadron fake background, we need to 
measure precisely the effective excess muon to electron fake rate ratio to derive the correct
background level.
The $D^*$ decays provide us with well-tagged kaon and pion samples. 
In our previous publication, 
the uncertainty in the fake rate value dominated the systematic errors. 
One major improvement of the current analysis is the better determination
of these  rates for muons and electrons from much 
larger tagged data samples obtained by using new data and adding more channels.

In this
analysis, in addition to the decay sequence $ D^{*+} \to D^o\pi^+ \to (K^-\pi^+)
\pi^+$, we also include  $D^{*+} \to D^o\pi^+ \to 
(K^-\rho^+)\pi^+$, and $D^{*o} \to D^o\pi^o \to (K^-\pi^+)\pi^o$ to get as many
events as possible. 
$K_S \to \pi^+\pi^-$ samples are also used to determine the pion fake rate
and are combined with the $D^*$ results to get better statistics. 
 Over 10,000 events were collected 
with either a $\pi$ or $K$ with momentum greater than 2.4 GeV from the
above channels.

\begin{figure}[b]
\centerline{\epsfig{figure=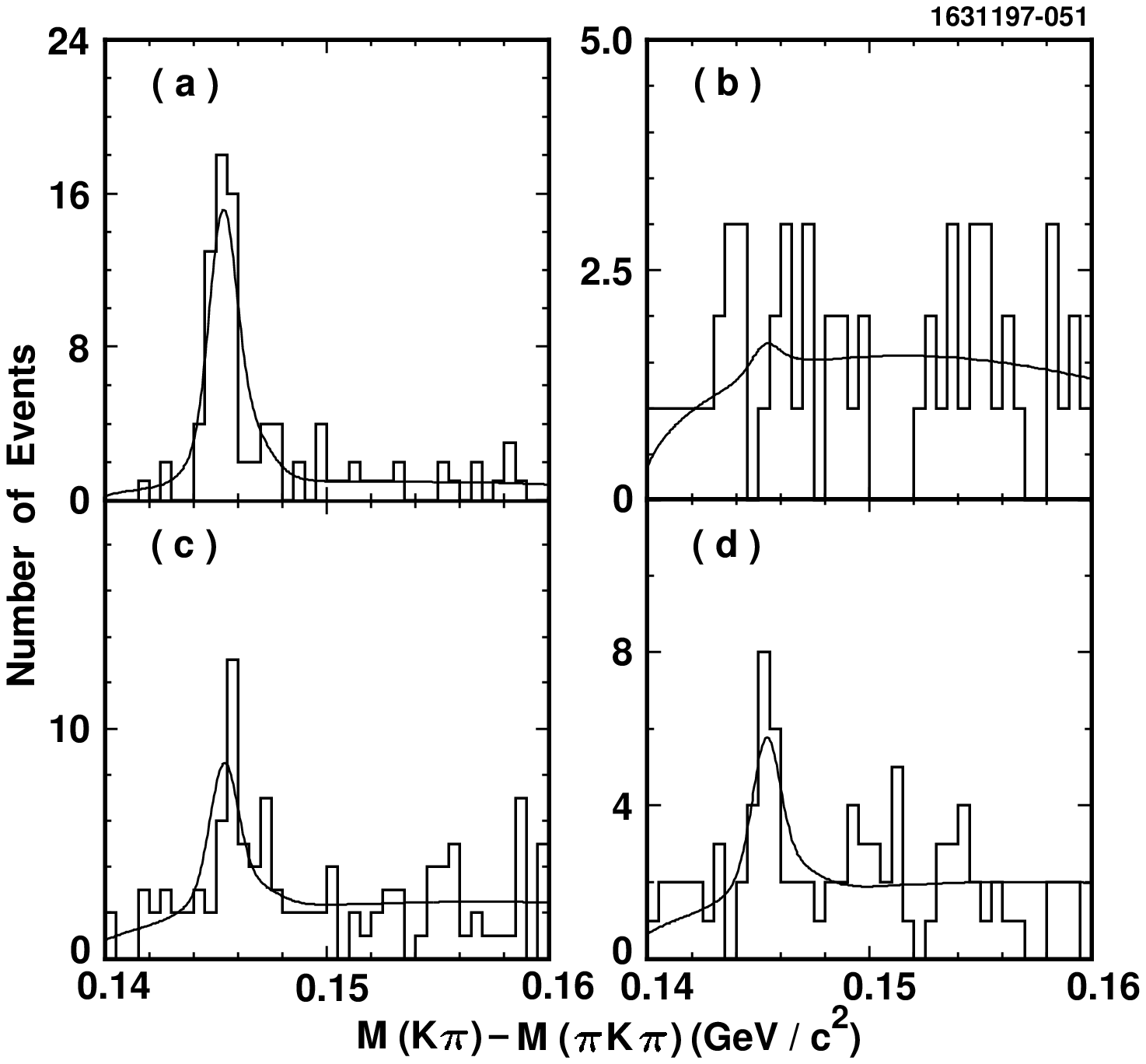,height=6in}}
\caption{\label{Fakeplot}  $M(\pi K\pi) - M(K\pi)$ mass difference
distributions for four cases of hadrons identified as leptons:
(a) kaon as muon, (b) kaon as 
electron, (c) pion as muon, (d) pion as electron. The signal shapes were
determined from the distribution of mass difference for
fully reconstructed $D^{*+}$ candiates. The area of the Gaussian component
and the normalization of the background are allowed to float.}
\end{figure}

  In Fig.~\ref{Fakeplot}  we show the $M(\pi^+ K^-\pi^+) - M(K^-\pi^+)$ mass difference after a cut on $K\pi$ mass consistent with  the
$D^o$ mass for kaons or pions which pass our cuts for
muons or electrons  The number of events is determined
by a fit with a double Gaussian for the signal and half-integer power polynomials 
for background. Both fitting function shapes are derived from the mass difference 
distribution without lepton identification suppression.
Our extracted fake rates (before decay in flight correction) are listed in
Table~\ref{table:Kpfake}. 
The same reconstruction methods are
used to collect kaon and pion samples from the channels
$D^o \to K^- \rho^+$, and $D^{*o} \to D^o \pi^o$ where $D^o$
goes to $K^-\pi^+$ sample. The fake rates are determined by fitting the  mass distributions
for the amount of signal. The fake rates derived from the different channels
are  summarized in Table~\ref{table:Kpfake}, and the 
weighted average fake rates are also shown.

\begin{table}[hbt]
\centering
\nobreak
\caption{Fake Rates for P $>$ 2.4 GeV/{\it c}}
\label{table:Kpfake}
\vspace*{2mm}
\begin{tabular}{l|cc|cccc}
Data Samples &  $\#~of$ & $\#~of$ & 
\multicolumn{4}{c}{Fake Rates (\%)} \\ 
&  $K$ & $\pi$ & $K/\mu$ & $K/e$ & $\pi/\mu$ & $\pi/e$ \\ \hline
$ D^{*+}(D^o \to K^-\pi^+)$ &   ~9404 & ~7461 & 0.94$\pm$0.11 & 0.04$\pm$0.05 &
	0.60$\pm$0.12 & 0.24$\pm$0.06 \\ 
$ D^{*+}(D^o \to K^-\rho^+)$ & ~1368 & ~~682 & 1.23$\pm$0.33 & 0.22$\pm$0.20 &
	0.30$\pm$0.40 & 0.15$\pm$0.21 \\ 
$ D^{*o}(D^o \to K^-\pi^+)$ &  ~3174 & ~2048 & 1.07$\pm$0.21 &  0.17$\pm$0.10 &
	0.84$\pm$0.35 & 0.60$\pm$0.31 \\ 
$ K_S \to \pi^+ \pi^- $ & - & ~3527 & - & - & 0.74$\pm$0.15 & 0.37$\pm$0.10 
      \\ \hline
Total/Average &   13964 & 13718 & 0.98$\pm$0.08 & 0.12$\pm$0.05 &
	0.65$\pm$0.08 & 0.31$\pm$0.06 \\  
\end{tabular}
\end{table}

 The contributions to the lepton fake rates
from kaon and pion decays in flight are not necessarily
included in the above procedure because particles decaying
close to the production point may not appear in the $D^o$ mass peak.
To account for this effect, we used a Monte Carlo
study of 200,000 $D^{*+} \to D^o\pi^+ \to (K^-\pi^+)\pi^+$ events.
After muon identification cuts are applied, the $D^{*+}$ mass difference
plot has a peak region used to derive the fake signal and a
tail away from the peak, which is due to events in which
the kaon decays.  We extract a correction factor to the
fake rate of {1.18$\pm$0.06} by
computing the ratio of the tail area to the peak area.
We find no events out of the $D^*$ mass difference peak
in which the pion has decayed.  This is because of the relatively
long pion lifetime and because the muon
momentum is very close to that of the parent pion.
  
We determine the hadron induced muon and electron fake background contributions
by multiplying the $\Delta M$ distribution of all tracks, excluding leptons,
by the effective fakes rates determined above. The fractions of kaon, pions
and protons are 67\%, 20\% and 13\% as ascertained from Monte Carlo simulation. The effective
Fake rates from protons and anti-protons are small, $\approx$0.1\%, and
almost equal for muons and electrons. 
\section{Results}
\label{Result}

\begin{figure}[b]
\vspace{-1.5cm}
\vspace{1cm}
\centerline{\epsfig{figure=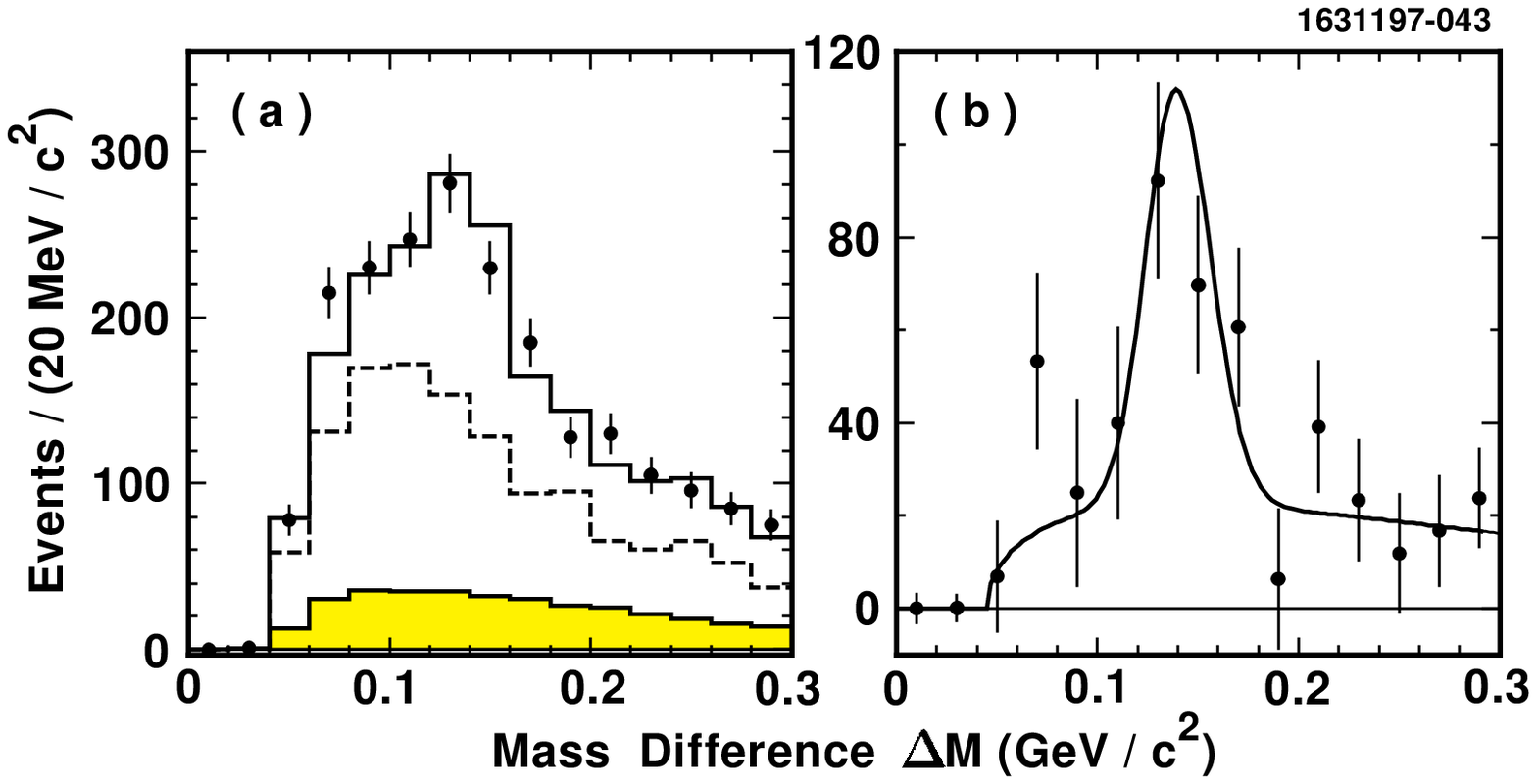,width=6.5in}}
\caption{\label{Fdata} (a) The $\Delta M$ mass difference distribution for 
$D_s^{*+}$
candidates for both the muon data (solid points), the electron data (dashed
histogram) and the excess of muon fakes over electron fakes (shaded). The histogram is
the result of the fit described in the text.
(b) The $\Delta M$ mass difference distribution for $D_s^{*+}$
candidates with electrons and excess muon fakes subtracted. The curve is a fit
to the signal shape described in the text.}
\end{figure}

The $\Delta M$ distributions for the muon and electron data and the calculated
effective excess of muon fakes over electron fakes are given in
Fig.~\ref{Fdata}(a). The histogram is the result of a $\chi^2$ fit of the muon
spectrum to the sum of three contributions: the signal distribution evaluated
with the $D_s^{*+}$ Monte Carlo plus random photon background evaluated with
the $D^{*+}$ sample, the scaled electrons, and the excess
of muon over electron fakes. Here, the sizes of the electron and fake
contributions are fixed and only the signal normalization is allowed to vary.

We remind the reader that the signal consists of two components, whose relative
normalization is fixed. These two components are the decay $D_s^{*+}\to\gamma
D_s^+$, $D_s^+\to\mu^+\nu$ and the direct decays $D_s^+\to\mu\nu$ and
$D^+\to\mu^+\nu$. Our measurement of the $D_s^{*+}/D_s^+$ production ratio
allows us to constrain the relative normalization. We find a signal of
182$\pm$22 events in the peak which are attributed to the process
$D_s^{*+}\to\gamma D_s^+$, $D_s^+\to\mu^+\nu$. We also find 250$\pm$38 events
in the flat part of the distribution corresponding to $D_s^+\to\mu^+\nu$ or
$D^+\to\mu^+\nu$ decays coupled with a random photon. The contribution  of a
real $D^+\to\mu^+\nu$ decay with random photons is not entirely negligible
since the $D^{*+}\to\gamma D^+$ branching ratio does not enter. The $D^+$
fraction is estimated to be about (18$\pm$8)\% relative to the total
$D_s^+\to\mu^+\nu$ plus random photon contribution.

To explicitly display the signal, we show, in Fig.~\ref{Fdata}(b), the 
$\Delta M$
distribution after the electrons and the fakes are subtracted.
The curve is a fit of
the data in Fig.~\ref{Fdata}(a)
 to the signal shape calculated from the $D_s^{*+}$ sample and 
random photon background calculated from the $D^{*+}$ sample. All of the
events in this plot are signal, the background having already been subtracted.

Using the fit result of  182$\pm$22 events,
we extract a width for $D_s^+\to\mu^+\nu$ by normalizing to the
the efficiency corrected number of fully reconstructed
$D_s^{*+}\to\gamma D_s^+$, $D_s^+\to\phi\pi^+$ events, 24740$\pm$1200$\pm$810
\cite {effcom}.
The efficiency for reconstructing the $\phi\pi^+$ decay is obtained
from Monte Carlo. We find
\begin{equation}
{\Gamma(D_s^+\to\mu^+\nu) \over \Gamma(D_s^+\to\phi\pi^+)} =
0.173\pm 0.023 \pm 0.035~~~,
\end{equation}
where the first error is the statistical error on the measured numbers of
$\mu^+\nu$ and $\phi\pi^+$ events.
The second error is the total systematic error of 18\%, whose
components are summarized in Table~\ref{table:TSystematics}.

\begin{table}[hbt]
\centering
\nobreak
\caption{Systematic Errors on Width Ratio (\%)}
\label{table:TSystematics}
\vspace*{2mm}
\begin{tabular}{lcr}
Source of Error &~~~Value&Size of error (\%)\\ \hline
Muon fake rate &(0.69$\pm$0.05)\% & 9 \\
Electron fake rate &(0.21$\pm$0.03)\% & 7 \\
$\pi/K/p$ fractions (sources of fakes) & 67\%/20\%/13\% & 7\\
$\mu$/e normalization & 1.01$\pm$0.03 & 9 \\
Detection efficiency & (4.2$\pm$0.3)\%&7\\
$D_s^{*+}/D_s^+$ production ratio &1.08$\pm$0.13& 8\\
$\phi\pi^+$ normalization&24740$\pm$1200$\pm$810    &3\\ \hline
Total systematic error &  &20  \\ 
\end{tabular}
\end{table}

The errors that arise from the relative muon to electron normalization, the
muon fake rate, the electron fake rate, and the $D_s^{*+}/D_s^+$ production
ratio, are estimated by fitting the data with each parameter changed by $\pm$
1$\sigma$. The error on the relative fractions of pions, kaons and protons
entering into the fake rate calculation is computed by changing the fractions
to 70\%, 20\% and 10\%, respectively. We judge this to be the outer limit at
90\% confidence level of the change possible in these ratios.
 This, in turn,  changes the excess muon to electron to fake rate by 12\%
 leading to a 7\% change in the yield.
A systematic error of $\pm$3\% for the detection efficiency of the
normalization mode $\phi\pi^+$ is also included.

The radiative decay rates for $D_s^+\to\ell^+\nu\gamma$ and $B^+\to\ell^+\nu\gamma$ 
have been considered by Burdman, Goldman and Wyler \cite{BGW}. They predict that
\begin{equation}
{\Gamma(D_s^+\to\mu^+\nu\gamma)\over \Gamma(D_s^+\to\mu^+\nu)}
= (1-10)\times 10^{-2}\mu^2_V {\rm ~GeV}^2,
\end{equation}
where $\mu^2_V$ is a vector coupling constant which has a value of
approximately 0.1 for the $D_s^+$ meson. While the radiative decay rate for
$B^+$ is  comparable to  the non-radiative rate, the radiative decay rate for
$D_s^+$ is estimated to be between  0.1\% and 1\% of the
non-radiative rate. Furthermore,  they also predict  that the radiative muon
and electron rates are equal,  so our electron  subtraction would remove any
residual effect.

\section{Conclusions}
\label{Conclusion}
 We have  measured the ratio of decay widths
\begin{equation}
\Gamma (D_s^+ \to \mu^+ \nu)/\Gamma (D_s^+ \to \phi\pi^+) =
0.173\pm 0.023 \pm 0.035~~~.
\end{equation}
To extract the decay constant $f_{D_s}$
we need to known the partial width for the $D_s^+\to\phi\pi^+$ decay.
The total $D_s^+$ width is well known because of
precise lifetime measurements \cite{PDG}, but the absolute $\phi\pi^+$
branching ratio has a large error.
Using the latest PDG average value ${\cal B}(D_s^+\to\phi\pi^+)$ of (3.6$\pm$0.9)\%,
and $\tau_{Ds}=(4.67 \pm 0.17) \times 10^{-13}$s, we find
\begin{equation}
f_{D_s} = (280\pm 19 \pm 28 \pm 34) {\rm ~MeV}~~~.
\end{equation}
The first error is statistical, and the second is systematic resulting
from our relative width ratio measurement, and the third
error reflects the uncertainty in the absolute $D_s^+\to\phi\pi^+$ branching
ratio. This result supersedes our previous one, using a data sample
that includes the one used in the previous analysis.
The reduction in the central value is primarily due to
the better measurement of the lepton fake rates that lowered the
pion/electron fake rate.

For comparison, we list in  Table~\ref{table:other} the old CLEO result and
published results from other experiments that used the $D_s^+\to \mu^+\nu$
decay to measure $f_{D_s}$.  We have changed the values of $f_{D_s}$ according
to the new PDG $D_s$ decay branching fractions for the normalization modes, and
have corrected the old CLEO result by using the new fake rates determined in
this analysis \cite{Bullshit}. The lowering of the central value of the old CLEO
result is mostly due to the change in the fake rate determination, which is now
much more precise.

\begin{table}[hbt]
\centering
\nobreak
\caption{Current Experimental Results on $f_{D_s}$ Using $D_s^+\to \mu^+\nu$}
\label{table:other}
\vspace*{2mm}
\begin{tabular}{lccc}
Collaboration & Observed & Published $f_{D_s}$  &
	Corrected $f_{D_s}$ \\
& Events & value (MeV) & value (MeV) \\ \hline
CLEO (old) \cite{cleo} & 39$\pm$8 & $344\pm 37 \pm 52 \pm 42$ &
	 $282 \pm 30 \pm 43 \pm 34$  \\ 
WA75 \cite{Bullshit} & 6 & $232 \pm 45 \pm 20 \pm 48$ &
	 $ 238 \pm 47 \pm 21 \pm 48$  \\

BES \cite{Bes} & 3 & $430 ^{+150}_{-130} \pm 40$ & Same \\ 
E653 \cite{E653} & $23.2\pm 6.0 ^{+1.0}_{-0.9}$ & $194\pm 35\pm 20\pm 14$ & $190\pm 34\pm 20\pm 
26$\\
CLEO (this work) & 182$\pm$22 & - &
	 $280 \pm 19 \pm 28 \pm 34$  \\ 
\end{tabular}
\end{table}
In addition, there are new results using the $D_s^+\to \tau^+\nu$
decay from the L3 collaboration \cite{L3} of $309\pm 58\pm 33 \pm 38$ MeV, and 
$330 \pm 95$ from the DELPHI collaboration \cite{DELPHI}. Our new measurement
gives the most accurate of $f_{D_s}$.

Theoretical predictions of $f_{D_s}$ have been made using many
methods.  Recent lattice gauge
calculations \cite{Lattices} 
give central values of 199 to 221 MeV with quoted errors in
the $\pm$40 MeV range. Other theoretical estimates
 use potential models whose values
\cite{Potentials} range from 210 to 356 MeV, and
QCD sum rule estimates \cite{QCDs} that are between 200 and 290 MeV.
Predictions for $f_{D_s}$ have also been made by combining theory with
experimental input.
Assuming  factorization for $\overline{B}\to D^*D_s^-$ decays
combined with measured branching ratios, gives a value of 
$f_{D_s}$ range of about 280 MeV with an error of about 60 MeV\cite{fdsfact}.
Use of experimental data on isospin mass splittings in the $D^*$ and $D$ system
gives a value for $f_D$ of 290 MeV \cite{fdiso}. ($f_{D_s}$ is thought
to be 10\% to 20\% higher than $f_D$).


\vskip 6mm
\centerline{\bf ACKNOWLEDGMENTS}
\smallskip

We thank Franz Muheim for his
creative efforts in the original analysis, and his continuing helpful
comments.
We gratefully acknowledge the effort of the CESR staff in providing us with
excellent luminosity and running conditions. We thank J. Cumalat and
W. Johns for interesting discussions.
J.P. Alexander, J.R. Patterson, and I.P.J. Shipsey thank 
the NYI program of the NSF, 
M. Selen thanks the PFF program of the NSF, 
G. Eigen thanks the Heisenberg Foundation, 
K.K. Gan, M. Selen, H.N. Nelson, T. Skwarnicki, and H. Yamamoto thank the 
OJI program of DOE, 
J.R. Patterson, K. Honscheid, M. Selen and V. Sharma 
thank the A.P. Sloan Foundation, 
M. Selen thanks Research Corporation, 
and S. von Dombrowski thanks the Swiss National Science Foundation 
for support. 
This work was supported by the National Science Foundation, the 
U.S. Department of Energy, and the Natural Sciences and Engineering 
Research Council of Canada.



\begin{thebibliography}{[9999]}
\setlength{\baselineskip}{2.0ex}
\bibitem{Formula1}
   J. L. Rosner, in {\bf Particles and Fields 3}, Proceed. of the 1988 Banff
Summer Inst., Banff, Alberta, Canada, ed. by A. N. Kamal and F. C. Khanna,
World Scientific, Singapore, 1989, 395.
\bibitem{chargeconj}Whenever a specific reaction or final state is mentioned,
consideration of the charge conjugate reaction or final state is also implied.
\bibitem{PDG}
   R.M. Barnett {\it et al.}, (Particle Data Group),
   Phys. Rev. D {\bf 54}, 1 (1996).
\bibitem{cleo}
   D.~Acosta {\it et al.}, {\rm Phys. Rev.} {\bf D49}, 5690 (1994).
\bibitem{Bullshit}
    S. Aoki {\it et al.}, Progress of Theoretical Physics {\bf 89}, 131 (1993).
    The WA75 value was based on the 1992 PDG value of
${\cal B}$($D_s^+ \to K^+K^-\pi^+)=(3.9\pm0.4)\%$ and ${\cal B}$($D^o \to \mu 
\nu X)=(8.8\pm2.5)\%$. We scale the WA75 result
to be $f_{D_s} = (238\pm 47\pm 21\pm 43)$MeV using updated PDG \cite{PDG} values of  
${\cal B}$($D_s^+ \to K^+K^-\pi^+)=(4.8\pm0.7)\%$ and
 ${\cal B}$($D^o \to e\nu X)=(7.7\pm1.2)\%$  which we use 
for ${\cal B}$($D^o \to \mu \nu X)$, after reducing the value by 3\% to
account for the smaller muon phase space.
\bibitem{Bes}
   J. Z. Bai {\it et al.}, Phys. Rev. Lett. {\bf 74}, 4599 (1995).
\bibitem{E653} 
  K. Kodama {\it et al.}, Phys. Lett. {\bf B382} 299 (1996).
\bibitem{CIId}
     Y. Kubota {\it et al.}, (CLEO Collaboration),
     Nucl. Instrum.  Methods Phys. Res., Sec. A {\bf 320}, 66 (1992).
\bibitem{Randomg}
M. S. Alam {\it et al.} (CLEO Collaboration), ``The Observation of the Radiative 
Decay $D^{*+}\to D^+\gamma$,"  CONF 96-7, ICHEP-96 PA01-80 (1996).
\bibitem{rocky}D. Atwood and W. Marciano, {\rm Phys. Rev.} 
{\bf D41}, 1736 (1990). 
\bibitem{Dstarplus}
To use our
$D^{*+}\to \pi^+ D^o$, $D^o\to K^-\pi^+$ events here, 
we eliminate
the measurements of the fast $\pi^+$ from the $D^o$ decay from both the
tracking chambers and calorimeter to simulate the neutrino, and call
the $K^-$ a muon.
\bibitem{denom}
The event sample used is the fully reconstructed sample of $D^{*o}$ events,
$D^{*o}\to \gamma D^o$, $D^o\to K^-\pi^+$, with the $K^-$ required to have
momentum above 2.4 GeV/c.
 
\bibitem{correff}
To compare with the efficiency in the $D_s^+\to\mu^+\nu$ channel, the muon
identification effiency of 0.85 must be taken into account, yielding
an overall efficiency of 4.1\%.
\bibitem{effcom}
The efficiencies quoted here are both for $\phi\pi^+$ momenta above 2.4 GeV/c.
To calculate the $D_s^{*+}\to\gamma D_s^+$, $D_s^+\to \phi\pi^+$ yield above
2.4 GeV/c $D_s^{*+}$ momentum, we remove the cut on the $\phi\pi^+$ momentum
and impose a cut on the $D_s^{*+}$ momentum.
\bibitem{goodVP}
The CLEO Collaboration is preparing a publication on the measurement of
the vector/pseudoscalar ratio in charmed-strange mesons.

\bibitem{L3}
M. Acciarri {\it et al.} (L3 Collaboration), 
 Phys. Lett. B {\bf 396}, 327 (1997).
\bibitem{DELPHI}F. Parodi {\it et al.} (DELPHI Collaboration),
 ``Measurement of the Branching Fraction
$D_s^+\to\tau^+\nu_{\tau}$," DELPHI 97-105 CONF 87, submitted to HEP'97
Conference Jerusalem, August 1997, paper 455.
\bibitem{Lattices}
See C. Bernard,``Lattice Calculations of Decay Constants," 
review talk presented at the
Seventh International Symposium on Heavy Flavor Physics,
 Santa Barbara, July 7-11, 1997, hep-ph/9709460, and references cited
 therein; A. X. El-Khadra {\it et al.}, ``$B$ and $D$ Meson Decay Constants
 in Lattice QCD," hep-ph/9711426 (1997). 
 
\bibitem{Potentials}
    H. Krasemann, Phys. Lett. B {\bf 96}, 397 (1980);
M. Suzuki, Phys. Lett. B {\bf 162}, 391 (1985);
S. Godfrey and N. Isgur, Phys. Rev. D {\bf 32}, 189 (1985);
S. N. Sinha, Phys. Lett. B {\bf 178}, 110 (1986);
P. Cea {\it et al.}, Phys. Lett. B {\bf 206}, 691 (1988);
S. Capstick and S. Godfrey, Phys. Rev. D {\bf 41}, 2856 (1988).
\bibitem{QCDs}
    C. Dominguez and N. Paver, Phys. Lett. B {\bf 197}, 423 (1987);
 S. Narison, Phys. Lett. B {\bf 198}, 104 (1987);
L. J. Reinders, Phys. Rev. D {\bf 38}, 947 (1988);
M. A. Shifman, Usp. Fiz. Nauk {\bf 151}, 193 (1987)
[Sov. Phys. Usp. {\bf 30}, 91 (1987)].
\bibitem{fdsfact}
    D. Bortoletto and S. Stone, Phys. Rev. Lett. {\bf 65}, 2951 (1990);
J. L. Rosner, Phys. Rev. D {\bf 42}, 3732 (1990);
H. Albrecht {\it et al.} (ARGUS Collaboration), Zeit. Phys. C {\bf 54}, 1 (1992);
T. Browder, K. Honscheid and S. Playfer, ``A Review of Hadronic and Rare
$B$ Decays," in {\bf  $\boldmath{B}$ Decays} 2nd edition, ed. S. Stone, World Scientific,
Singapore, p 158 (1994).
\bibitem{fdiso}
    J. F. Amundson {\it et al.}, Phys. Rev. D {\bf 47}, 3059 (1993).

\bibitem{BGW}
G. Burdman, T. Goldman and D. Wyler, Phys. Rev. D {\bf 51}, 111 (1995).


\end{thebibliography}
\end{document}